\documentclass[english, 12pt, titlepage]{article}

\usepackage{graphicx}
\usepackage[stretch=10]{microtype}
\usepackage{todonotes}

\usepackage[backend=bibtex, autocite=footnote,style=verbose-trad1]{biblatex}
\let\cite\autocite

\bibliography{STS.bib}

\begin{document}
\title{Man and Machine: Questions of Risk, Trust and Accountability in  Today's AI Technology}
\author{Piyush Ahuja\thanks{The author will graduate from IIT Delhi in 2013 with a major in Maths and Computing. His current research straddles the fields of Algorithms and Economics. The research presented here was carried out in his fourth year.}
\\Indian Institute of Technology Delhi}

\maketitle

\begin{abstract}
Artificial Intelligence began as a field probing some of the most fundamental questions of science - the nature of intelligence and the design of intelligent artifacts. But it has grown into a discipline that is deeply entwined with commerce and society. Today's AI technology, such as expert systems and intelligent assistants, pose some difficult questions of risk, trust and accountability. In this paper, we present these concerns, examining them in the context of  historical developments that have shaped the nature and direction of AI research.  We also suggest the exploration and further development of two paradigms, human intelligence-machine cooperation, and a sociological view of intelligence, which might help address some of these concerns.
\end{abstract}

\section{Introduction}

\begin{quote}
	\textit{``Synthesis defines an ambitious ‘put-a-man- on-the-moon’ goal. By doing so, it forces scientists and engineers to cross uncharted terrain in pursuit of the goal. This requires the solution of unscripted problems that are not normally encountered through either observation or analysis... synthesis drives the evolution of paradigms"}\autocite[It is hard to understate the role of synthesis in science and technological development. The above quote has been taken from the context of \textit{synthetic biology, see}][]{brenner2005synthetic}
\end{quote}

Artificial Intelligence (AI) is the field of computer science which aims to create, or synthesize, intelligence. The pursuit of creating intelligent machines has contributed not only to psychology, cognitive science, neurology and philosophy, but has also given birth to whole new branches of research. 

The field of AI research was formally founded at a conference on the campus of Dartmouth College in the summer of 1956\autocite[The conference was chaired by J.McCarthy, M. Minsky, N. Rochester and C.Shannon. For a full report, see][]{mccarthy2006proposal}. It has come a long way since then, having gone through many cycles of boom and bust,``AI winters" and summers\autocite{nilsson2010quest}. Now AI applications have left the annals of Department of Defense R\&D, and  trickled down to everyday use. They can be found in common place consumer items and inexpensive intelligent toys, even though very often consumers fail to recognize the technology source.  For example, the Kinect, which provides a 3D bodymotion interface for the Xbox 360, uses algorithms that emerged from lengthy AI research\autocite{smisek20133d, khoshelham2012accuracy, oikonomidis2011efficient}.

The adoption and integration of AI based technology in all spheres has followed a pattern reminiscent of many modern technologies, including, for example, the internet, digital computing\autocite[The reader is referred to James Cortada's essay in Technology and Culture's April 2013 issue.][]{cortada2013new}, and mobile telephony. The pace has been such that historians and sociologists have barely had an opportunity to study its arrival, use and implications for society. Most discussions in this direction have been based on a priori and futuristic assessments,  discussing the future of AI technology and what it might eventually turn out like.  For instance, researchers have raised concerns about AI machines becoming malicious (unfriendly), apathetic or uncontrollable, often evoking images of killer robots and future wars between man and machine\autocite{joy2000future, bostrom2008global, yudkowsky2008artificial, bostrom2002existential, bostrom2003ethical}. The importance of incorporating `friendliness' in AI research is repeatedly stressed in these discussions.  Some have focussed on the concept of `intelligence explosion', forewarning a future event where radically self-improving machines reach a state where it is impossible to predict or comprehend their actions\autocite{muehlhauser2012intelligence}.  Others have noted that increasing dependence on decision-making intelligent machines may itself lead to a world where it is impossible for humans to survive without them, leaving machines effectively in control\autocite[In his now famous article ``Why The Future Doesn't Need Us", computer scientist Bill Joy quotes Ted Kaczynski, ``As society and the problems that face it become more and more complex and machines become more and more intelligent, people will let machines make more of their decisions for them, simply because machine-made decisions will bring better results than man-made ones. Eventually a stage may be reached at which the decisions necessary to keep the system running will be so complex that human beings will be incapable of making them intelligently.' See ][]{joy2000future, kaczynski2005unabomber}. 

Such discussions, though very important in their own right, tend to  take the attention away from issues of risk and trust posed by AI-based technology which has already diffused in the society.  In this essay, we argue that current AI technology (e.g., expert systems and intelligent assistants) is based on a notion of intelligence which is somewhat different from the notion of`general intelligence' in popular perception.  The intelligence or expertise of these machines is measured through their performance in certain specialized contexts, and little else. Further, this confusion of what AI technology is based on, and how it works, itself has risk and trust-related consequences. As AI researcher Elizier Yudkowsky had noted,  ``By far the greatest danger with Artifical Intelligence is that people conclude too early that they understand it"\autocite{yudkowsky2008artificial}.

We start off this essay by tracing the history of current AI research in Section 2. Various factors which have steered AI research in a particular direction, and diminished in others, have been outlined. Section 3 examines the nature of AI systems today, illustrating, through the examples of expert systems and intelligent assistants, the questions of risk and trust they pose to society. In Section 4, we suggest two paradigms, which might complement basic research in AI to address some of these concerns . The discussion is summarised in the conclusion.

\section{Historical Developments}

Ever since AI’s inception, there has been no clear consensus on what constitutes `intelligence'. The subject has drawn from a broad array of disciplines - Philosophy, Logic, Biology, Psychology, Statistics and Engineering. In the absence of an agreed-on curriculum for training students in AI,\footcite{nilsson2010quest} new researchers who enter the field bring with them different standards, traditions and problems. As a result, one big challenge that AI has faced is that the research effort has been characterized by a multiplicity of approaches, each endeavoring to attain some specific objective . This multigoal character of AI research has crystallized into its theoretical pluralism, and its institutionalization by means of competing groups with different aims\footcite[Fleck, 1982: 172 in]{schwartz1989artificial}. It has been a major factor in the eventual branching of Artificial Intelligence to various subfields, e.g., Knowledge Representation, Machine Learning and Natural Language Processing.

%Ever since AI’s inception, there has been no clear consensus on what constitutes `intelligence'.  The subject has drawn people from a broad array of disciplines - Philosophy, Logic, Biology, Psychology, Statistics and Engineering, having different degrees of inclusion in various technological frames\autocite[The concept of technological frames appears in the social connstruction of artifacts, referring to stability in``ways of thinking" and ``fixed patterns of interaction" that ``emerge around them". ``People with a high degree of inclusion in a technological frame will find it difficult to imagine other ways of dealing with the world, of using these things radically differently or even not using them at all", see][]{pinch1987social, bijker2001understanding, bijker2007dikes, orlikowski1994technological}. As a result, one big challenge that AI has faced is that people bring with them different standards, traditions and problems. Historically, there has been no agreed-on curriculum for training students in AI, and new researchers entered the field from various universities with quite differing points of view. 
Up until the early 1970s, AI researchers dealt with highly theoretical problems probing the nature of intelligence, and the researchers pursued projects that were staged in highly controlled laboratory settings\cite[p 265]{nilsson2010quest}. The heavy initial funding (by DARPA) into such pathbreaking research was fuelled partly due to the cold war\cite[p. 66]{whitby1996reflections}, and partly due to the highly optimistic claims made by its pioneers and their early theoretical successes. 

However, such theoretical approaches to AI gradually came to be thought off by many computer scientists as fringe activities that did not adhere to rigorous scientific standards - some even viewed AI as `a field that housed charlatans'\cite[p 339]{nilsson2010quest}. Niel Nilsson, one of the founding fathers of the discipline, recalls that when he first interviewed for a position at SRI in 1961,\footnote{Stanford Research Institute} a researcher had warned him against joining research on neural networks because it was `premature', and his involvement with it could damage his reputation. This concern for `respectability' had a stultifying effect on many AI researchers.\footnote{It is interesting to note that many researchers in AI today deliberately call their work by other names, such as informatics, knowledge-based systems, cognitive systems or computational intelligence. One of the reasons is that these names help to procure funding. As New York Times\cite{markoff2005behind} reported in 2005, ``Computer scientists and software engineers avoided the term artificial intelligence for fear of being viewed as wild-eyed dreamers"}

In 1969, the Mansfield Amendment dealt another significant blow to the field. The Amendment put DARPA under increasing pressure to fund only `mission-oriented direct research, rather than basic undirected research'\autocite[][under "Shift to Applied Research Increases Investment" (only the sections before 1980 apply to the current discussion)]{national1999funding}. The creative, freewheeling exploration that was characteristic of the early pioneering work in AI gradually came to be viewed as a burden. Instead, the money was directed at specific projects with clear objectives, such as autonomous tanks and battle management systems\cite[under ``Shift to Applied Research Increases Investment"]{national1999funding}. Not only did this greatly influence the direction of research, but also, perhaps more importantly, swayed the spirit that guided work in the field. This has been noted by Marvin Minsky\cite{roush2006marvin}, co-founder of MIT's AI Lab and a leading cognitive scientist:
\begin{quotation}
\textit{	``In the early days, DARPA supported people rather than proposals. There was a lot of progress from starting in 1963 for about ten years in all branches and all approaches aimed to modeling intelligence. But the Mansfield Agreement made it much harder to support visionary researchers. At the same time, the American corporate research community started to disappear in the early 1970s. Bell Labs and RCA and the others essentially disappeared from this sort of activity."}
\end{quotation}

Also, by the end this period, the power of AI methods had already increased to the point where realistic applications seemed within reach\cite[p 265]{nilsson2010quest}. This gave rise to what Minsky calls `the entrepreneur bug'\cite{roush2006marvin}. He attributes the disappearance of young scientists in that period to an increased tendency to patent things, start start-ups and make new products. Support for original theoretical research in areas like commonsense reasoning eventually fizzled out.

The collective consequence of all these factors was that by the end of the 70s many had people diverted to highly specialized subfields that solved specific real-world problems. These problems ranged from Speech Recognition and Understanding Systems, Consulting Systems, Understanding Queries and Signals to Computer Vision\cite[p 265]{nilsson2010quest}. Since these also had commercial applications, funding and notions of academic respectability became directly associated with these specialized problems.

Most AI researchers around this period eventually adopted the premise that general human-like intelligence can be developed by combining the programs that solve various subproblems using an integrated agent architecture. This was the Intelligent Agent paradigm, an approach that had become widely accepted in the AI community by the end of the 90s.\footcites[][pp. 27, 32–58, 968–972]{poole1998computational}[][pp. 7–21] {luger2005artificial}[][pp. 235–240]{russell2003artificial}

An intelligent agent is a system that perceives its environment and takes actions which maximize its chances of success\autocite[The definition used here is due to Russell and Norvig, see][Other definitions also include knowledge and learning as additional criteria.]{russell2003artificial}. The simplest intelligent agents are programs that solve specific problems. The paradigm gave researchers license to study isolated problems and find solutions with actual applications, without agreeing on one single approach.\footnote{An agent that solves a specific problem can use any approach that works – some agents are symbolic and logical, some are sub-symbolic neural networks and others may use new approaches.} For perhaps the first time in AI, a partial consensus on a notion of intelligence was achieved, since it was well aligned with the pluralistic character of AI, the constraints of funding, the notion of academic respectability, as well as relevant to industrial applications. Many fundamental questions related to intelligence - symbolic and commonsense reasoning, intuition, imagination,  creativity, and emotional intelligence, were left answered and unexplored.
%These differences have stimulated creative development,\footnote{For example, the Intelligent Agent paradigm, which involves a system modelling its environment and directing its actions so as to maximize its chances of success, came about from the marriage of an economist's definition of a rational agent, and a computer scientists' definition of an object or module.} but they have also worked against the maturation of AI as a serious science. 

\section{AI Today}

\begin{quote}
	\textit{``Today's AI is about new ways of connecting people to computers, people to knowledge, people to the physical world, and people to people"}. 
	- Patrick Winston, MIT AI Lab briefing, 1997
\end{quote}

\subsection{All That Glitters}

Today, with the increased power of relatively inexpensive computers, availability of large databases and growth of the World Wide Web, AI technology like \textit{expert systems} and \textit{intelligent assistants} have slowly diffused into our society. These have been said to heralding a new age and revolutionizing the way we live today. Such proclamations might be true, but not necessarily in the context in which they are made. Most people overlook the fact that most of these systems do not work on the premise of approximating human cognitive abilities, but on performing certain tasks in highly specialized scenarios. They do well in performing certain jobs that are complicated for people to do, but they are far from being capable of carrying out tasks that are simple for humans to do. Many elements, like common-sense reasoning, ethical judgements and decision-making under uncertainties, which might be core components of intelligence, are missing. The deployment of such systems and their celebration, besides diverting research from other fields, has social, ethical and risk - related consequences. 

Performance in specialized contexts (e.g., processing natural language, playing chess, recognizing patterns)  are the sole basis for intelligence for current AI based systems. There is a certain gap in what these systems are represented as doing, and what they actually do. This  can mislead people on two levels. One,  that they create a notion that they work in the same way that humans do is (human-experts in the case of expert systems and assistants in case of digital assistants). On a second level, they create the impression that these are free from prejudices, personal bias and errors that might affect a human counterpart. The actual implementation and mechanism - what goes behind an AI technology - is entirely invisible to the user. The ‘knowledge engineering’, as it is called in the case of expert systems, and the underlying architecture, might involve restrictions and selective representation of data, and unavoidable biases resulting from the organization of data and architecture of the machine. 

\subsection{Risk}

Another, possibly graver issue is that of handling of uncertainties by AI systems. Inductive knowledge of computer systems is inadequate. The number of combinations of possible inputs and internal states of a computer system of any complexity is huge, and especially in AI technologies, such complexity is intractable. Even with highly automated testing, it will seldom be feasible to exercise each and every state of a system to check for errors or underlying design faults, or bugs that may have caused them. As computer scientist Edsger Djikstra famously put it in 1969, ``Program Testing can be used to show the presence of bugs, but never to show their absence!" Because bugs may lurk for years before becoming manifest as a system failure, no guarantees can be offered.

In the case of expert systems, although they do reason usefully and economically about specific problems in medicine, geology, chemistry and other delimited areas, they are acknowledged to be brittle (that is, they break down) when confronted with problems outside their area of expertise or even on problems within their area of expertise, if knowledge were needed that had not been provided in their rulebooks. They don’t know what they don’t know, and therefore might provide wrong answers in cases where a human expert would do better. 

A colorful anecdote involving John McCarthy\footnote{John McCarthy was one of the founding fathers of the field. He purportedly coined the term `Artificial Intelligence'} illustrates this. In an interaction with the medical expert system MYCIN, he typed in some information about a hypothetical patient, saying that he was male and also saying that he underwent amniocentesis. MYCIN accepted all that without complaint! That male patients don’t get pregnant was not considered part of the `expert knowledge' that MYCIN needed to be given\cite[p 407]{nilsson2010quest}.

Donald Mitchie, a British researcher, examines this problem through his concept of the `human window' on the reasoning of the program, where its behaviour is `scrutable'. Outside of this window, Mitchie observes, it is impossible to tell whether the program is being exceedingly clever or is just malfunctioning\cite[Michie, 1984 in]{schwartz1989artificial}.

Reliance on experts is said to be an inevitable aspect of high modernity\cite{schwartz1989artificial}. As we enter an age of knowledge-intensive `information society', how we employ expert systems based on AI will turn out to be crucial in shaping our society.

The handicaps involving AI technology, as well as their associated risks, are greatly amplified in those systems that are critical to human safety or security. A disturbing example of this is provided by Donald MacKenzie\cite{mackenzie2004mechanizing}, a Professor of Sociology at the University of Edinburgh:

\begin{quotation}
	``\textit{On October 5, 1960, United States' Ballistic Missile Early Warning System (BMEWS) went off, indicating several missile launches from a general area in Siberia, and sent everyone in a panic mode. A level 5 alarm meant 99.9 percent likelihood that a missile attack had been launched. If that were true, ellipses should have been forming on the war room’s display map of North America in Colorado Springs and should have started to shrink, indicating the target of the attack. Yet no ellipses were being formed, and the “minutes-to-go” indicator showed nothing. Soviet Union premier Nikita Khrushchev was in New York, attending the General Assembly of the UN. Deputy General Slemon decided that Soviet Union was highly unlikely to attack the US while its leader was in New York. It was characteristic human reasoning. He also knew that the BMEWS had been operational for only four days and was still being 'run-in.' So, no action was eventually taken. It was later found out that the powerful radars in Greenland, designed to detect objects upto 3,000 miles distant, were fooled by the reflections from the slow-rising moon over Norway. The BMEWS intepreted the radar echoes as sightings of multiple objects, and the engineers had left this phenomenon unaccounted for while designing the system.}"
\end{quotation}

Since expert systems are deemed to emulate the decision-making ability of a human expert, and intelligent assistants work autonomously, human beings are increasingly being decoupled from the inner workings and implementations of such technology. In the case of the 1960 nuclear false alarm, human beings had remained in the loop, and common-sense had prevailed. With the increasing deployment of AI technology everywhere, it might not always be the case.

In his conception of the `risk society', contrasting the nineteenth-century and the present, German sociologist Ulrich Beck\cite[p. 21]{beck1992risk} had noted, ``hazards in those days assaulted the nose or the eyes and were thus perceptible to the senses, while the risk of civilization today escapes perception." This idea, apart from its associations with nuclear technology and biotechnology,  manifests itself in the domain of AI, and should be read in its context as well. The current AI technologies pose dangers of an entirely different kind - involving invisible contingencies, the seriousness of which layperson’s eyes cannot judge.\footnote{The Y2K problem at the end of the previous millennium was an important indication of the kind of risk advanced computing technologies entail. The millennium bug episode highlights both the dependence of the modern societies upon computing, as well as the difficulty of forming a judgment of risks posed by that dependence.}

\subsection{Trust}

\framebox{\begin{minipage}[c][1\totalheight][t]{1\columnwidth}{Siri is a personal assistant application for iOS. The application uses natural language processing to answer questions, make recommendations, and perform actions by delegating requests to an expanding set of web services. The software also supposedly adapts to the user's individual preferences over time and personalizes results, as well as accomplishing tasks such as making dinner reservations and reserving a cab. It is a spin-out from the SRI International Artificial Intelligence Center, and is an offshoot of the DARPA-funded CALO project, described as the largest artificial-intelligence project ever launched. The project brings together experts in machine learning, natural language processing, and knowledge representation, human-computer interaction, flexible planning and behavioral studies. The CALO software learns by interacting with and being advised by its users and is meant to help users with military decision making task.} \end{minipage}}
\vspace{5 mm}

Once technical objects are stabilised, they become instruments of knowledge. As already noted the current AI technologies, like expert systems or intelligent assistants, portray an impression of intelligence which is somewhat misleading. The actual engineering of these systems is hidden from the user, and in most cases their knowledge and utility arises from connections to databases and systems which are external to the environment in which they are deployed. Thus, the knowledge these technical systems gather can be `exported'\cite{akrich1992scription}. To take a mild example - the Apple iOS assistant application of Siri, which is a by-product of AI research, draws its knowledge from the Cloud and Internet databases\cite{Costa:2011:Online}. 

How does a user trust that the AI technology is working for his (her) welfare, and not some hidden agenda placed there, intentionally or inadvertently, by the programmer? How does the user trust that it will behave for his welfare, even in his absence, if the user doesn't have detailed knowledge of the system's underlying hardware and software limits, not to mention the subtle bugs. AI systems, being highly automated and self-sufficient (and having a deep, hidden architecture), raises trust issues which are qualitatively different from the `tool-like technologies' of the past. As more such AI technologies make their way to the general public, the convenience of using these systems and the deceptive nature of their intelligence will tend to replace the concerns over privacy and security.

\begin{figure}
	\centering
		\includegraphics[width=0.95\textwidth]{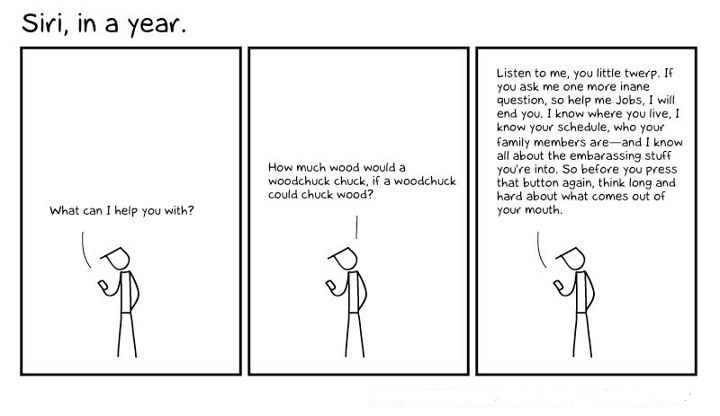}
	\label{fig:siri}
\end{figure}

\section{New Directions}

It has been shown that extraneous sociological factors have been responsible for the gradual deviation of AI research from its original objectives and its acceleration in one particular direction - which has probably led to the inadequacies presented by AI technologies of today. Basic areas like judgement under uncertainty,  intuition, commonsense reasoning, inventiveness, non-linear thinking have lagged behind, and should be encouraged. Going forward, this paper presents the case for two complementary methodologies that help in addressing the concerns arising out of the technologies and research as it is today.

\subsection{A Sociological View of Intelligence}

Almost all of the earlier models on which AI and cognitive science rested assumed that intelligence is not socially constituted and not socially situated. In fact , the traditional AI techniques hinged on the following assumptions\cite{Restivo:2001}:

\begin{itemize}
	\item Human mentality is a freestanding, individual, brain based phenomenon; and
	\item Human mentality is best understood in logical, linguistic, and rational terms
\end{itemize}
Most of the current AI technologies have resulted from the research grounded on such similar assumptions. But the success of these applications, especially of those in the form of intelligent assistants and expert systems, depends on the basis of how they perform in contexts and situations that are essential social and cultural in nature. Most of the inadequacies of these technologies highlighted above, result from their inability to handle practical social situations and uncertainties. 

The idea that social and cultural factors are not only important but primary is not yet a widely appreciated or understood possibility. In the past, Joseph Weizenbaum has observed that intelligence manifests itself only relative to specific social and cultural contexts \cite{clocksin2003artificial}. There is a need to develop a sound sociological basis for Artificial Intelligence. Social and Cultural assumptions should be incorporated in the worldviews guiding the work in AI. Intelligence, thus, should be understood as how well an entity performs in social situations, in addition to the inner architecture and cognitive model that brings about that behavior. A Sociology of Artificial Intelligence will result in more robust technologies in the future, which are better equipped to handle real world scenarios.

\subsection{Human-Augmented AI}

When Doug Engelbart was creating early computer interfaces and mapping systems, he firmly maintained his belief that the machine was meant to be an augmenter, not prosthesis. J.C.R. Licklider\cite{licklider1960man}, the computer science pioneer who had a profound effect on the development of technology and the Internet, also had the vision of enabling man and machine to cooperate in making decisions and controlling complex scenarios together, instead of compromising the flexibility by being dependent on predetermined programs. Future research should focus on similar principles. 

To get a flavour of the potential power of this concept, consider the 2005 freestyle chess tournament in which man and machine could enter together as partners, rather than adversaries. Initially, even a supercomputer was beaten by a grandmaster with a relatively weak laptop. But the real suprise which took everyone off-guard came at the end. The eventual winner was not a grandmaster with a supercomputer, but actually two American amateurs using three relatively weak laptops.  Their ability to effectively use their computers to deeply explore specific positions effectively counteracted the superior chess knowledge of the grandmasters and the superior computational power of other adversaries\cite{shyamshankharted}. This astonishing result, of average men, average machines beating the best man, the best machine, strinkingly illustrates the strength of human-machine cooperation - that the right symbiosis can be much more powerful than the sum of individual parts.

The power of Man and Machine working together has also manifested in a totally different but very relevant setting - Protein Folding. There are more ways of folding a protein than there are atoms in the universe. This is a world-changing problem with huge implications for our ability to understand and treat diseases. Supercomputers have traditionally struggled in this area. When computer scientists at created Foldit, a game where non-technical, non-biologist amateurs visually rearrange the structure of the protein (allowing the computer to manage the atomic forces and interactions and identify structural issues) , it was found that the players beat supercomputers 50 percent of the time and tied 30 percent of the time.  Recently, the structure of the Mason-Pfizer monkey virus, a protease that has eluded determination for over 10 years, was opened up to a group of online gamers (through Foldit), who competed to model the protein, with all the associated scores, points, and rankings of a game. The players finished the model within 10 days - a notable and major scientific discovery\cite{shyamshankharted}\cite{woods:2012:Online}.

Amazon partially taps into the principle of Human-Augment AI through a concept which they call `Artificial Artificial Intelligence'. The premise is simple: Since Humans outperform AI in many simple tasks (like recognizing faces or sorting patterns), so why not farm out computing tasks to people, instead of machines? AAI has been employed as part of The Amazon Mechanical Turk (MTurk), a crowdsourcing Internet marketplace that enables computer programmers (known as Requesters) to co-ordinate the use of human intelligence to perform tasks that computers are unable to do yet\cite{Bolt:2005:Online}.

Current AI technologies lack common sense and symbolic reasoning, inventiveness, non-linear apparoaches, iterative hypotheses - aspects of intelligence which come naturally and easily to humans. On the other hand, machines have been better and ever improving in handling scale, volume and computation. Thus, an approach which harnesses the best of both worlds holds great potential. In completely manual and completely automated system, users and machines are effectively decoupled and too often, systems fail because they are not designed as a whole system with people and machines working in harmony. The power lies of Human-Augmented AI will lie in expressing the `black box' and making it transparent. The human mind will enhance the machine’s solution by filling in the gaps.

A field of human intelligence - machine intelligence cooperation will thus help answer the concerns arising from the increasing separation of human beings from the inner workings and implementations of AI technology highlighted in the previous sections. This approach, traditionally under-appreciated and unexplored. For example, a language of machine-plus-human interaction has not yet been  developed. There is a need to revaluate and reframe the conventional `Man vs Machine' dichotomy towards a common `Man and Machine' framework\cite{gupta2012}. \footnote{The early signals indicating the potential of Human-Augmented AI technology can be read in the Big Data landscape. The idea of a `Man and Machine framework' to generate commercial insights has recently been pioneered by Opera Solutions, while software company Palantir Technologies takes on important real world problems (e.g. counterterrorism) using a concept called `Intelligence Augmentation'.} Research oriented at technologies which involve both humans and machines performing intelligent tasks in the context of an integrated system should be encouraged. This would involve designs where both humans and machines have resposibilities, require access to resources, and have particular knowledge appropriate to tasks. Tasks maybe performed in parallel, or may require results or permission from the other. The field of Human Intelligence-Machine Intelligence cooperation would define roles to each as tools or assistants. Different types of knowledge would have to be distinguished: designers, end users, and maintenance people for instance. Humans would thus be designed in the process, and friction between man and machine, minimized.

\section{Conclusions}

The state of AI research and technology, as it stands today, is not without blemishes. Under the shining surface, it hides some fairly serious concerns and issues. Most of these risk and trust related issues arise from an imbalance resulting from the impact of extraneous sociological factors on AI research. The exploration of two paradigms - Sociological AI and Human-Augmented AI - will go a long way in addressing these concerns, and hold great potential as a research agenda for the future. 

Besides this, it is essential that the funding agencies appreciate the importance of long-range basic research in AI. As Nils Nillson\cite{nilsson1983artificial} observes, AI, perhaps together with molecular genetics, will be society's predominant scientific endeavor for the rest of this century and well into the next - just as physics and chemistry predominated during the decades before and after 1900.

\nocite{*}   % All bibliography items appear without citation in the text

\printbibliography
\end{document}